\documentclass[prb,twocolumn,amsmath,amssymb,superscriptaddress]{revtex4}
\usepackage{graphicx}% Include figure files
\usepackage{dcolumn}% Align table columns on decimal point
\usepackage{bm}% bold math

\usepackage{float}

\usepackage{amsmath}

\usepackage{epstopdf}

\begin{document}

\title{Coupling of spin and lattice modes in the S=1/2 two-dimensional
antiferromagnet K$_{2}$V$_{3}$O$_{8}$ with magneto-dielectric couplings}
\author{K.-Y.~Choi}

\affiliation{Department of Physics, Chung-Ang University, 221 Huksuk-Dong, Dongjak-Gu,
Seoul 156-756, Republic of Korea}

\author{P.~Lemmens}

\affiliation{Institute for Physics of Condensed Matter, TU
Braunschweig, D-38106 Braunschweig, Germany}

\author{V. P. Gnezdilov}
\affiliation{B. I. Verkin Institute for Low Temperature
Physics NASU, 61164 Kharkov, Ukraine}

\author{B.~C.~Sales}
\affiliation{Oak Ridge National Laboratory, P.O. Box 2008,
Oak Ridge, Tennesse 37831, USA}

\author{M. D. Lumsden}
\affiliation{Oak Ridge National Laboratory, P.O. Box 2008,
Oak Ridge, Tennesse 37831, USA}
\date{\today}

\begin{abstract}
Lattice dynamics and magnetic excitations are investigated to elucidate the origin of magneto-dielectric effects in the S=1/2 two-dimensional quantum spin compound  K$_2$V$_3$O$_8$. We find evidence for lattice instabilities at 110~K and 60~K as optical phonon anomalies and a soft mode at 26~cm$^{-1}$ in A$_1$ symmetry. Two-magnon excitations in B$_1$ symmetry show an unconventional double-peak structure and temperature dependence.
This suggests the existence of a split mode near the zone boundary caused by a mixing of spin and lattice modes.
\end{abstract}


\maketitle

% \narrowtext

%\newpage

\section{INTRODUCTION}

In recent years, the cross-coupling effects of magnetic and
electric properties in a single material have been intensively investigated
due to the fascinating underlying physics and
the potential to realize multifunctional devices.~\cite{Cheong}
Prominent examples are found in multiferroic materials such as charge ordered
LuFe$_2$O$_4$, magnetically driven RMn$_2$O$_5$ (R=rare earth), and
lone pair multiferroic compounds R(Fe,Mn)O$_3$.~\cite{Wang,Kimura,Hur}
In such material classes, low-energy excitations are given by so-called electromagnons,
which are coupled spin and lattice excitations.~\cite{Kenzelmann}  However,
it turns out that electromagnons are not a monopoly of the multiferroic materials.
Electromagnons have also been observed in the paraelectric phase of a conical-spin magnetically ordered hexaferrite Ba$_2$Mg$_2$Fe$_{12}$O$_{22}$.~\cite{Kida}

Here the question is to what extent phonon-magnetic coupled modes
retain their hybrid nature in nonmultiferroic materials but with magnetodielectric couplings.
For such less strongly coupled materials, magnetism is not mutually coupled to ferroelectricity
but it is tied to dielectric properties. The A$_2$B$_3$O$_8$ fresnoites (A=K,Rb,Cs, NH$_4$;
B=Ti, V, Mn, Cu) are suitable for studying the coupling of a soft phonon mode to magnetic excitations because they are susceptible to displace structural phase transition and show
novel magnetic properties and nonlinear optical properties.~\cite{Withers,Hochte,lumsden,sales,Yuan}   Actually, magnetodielectric effects have been reported in K$_{2}$V$_{3}$O$_{8}$ and are attributed to the interplay between structural and electronic degrees of freedom.~\cite{Rai}

K$_{2}$V$_{3}$O$_{8}$ is the  S=1/2 square lattice antiferromagnet
which has a space group {\it P4bm} and
lattice parameters a=8.870~\AA\,  and c=5.215~\AA\,  at room temperature.~\cite{galy}
The vanadium layer consists of slabs of corner-sharing VO$_5$ square pyramids and
VO$_4$ tetrahedra, separated by nonmagnetic, interstitial K$^{+}$ ions.~\cite{liu} The
magnetic behavior is described by a Heisenberg model with an
exchange coupling constant of $J=12.8$~K  together with a
Dzyaloshinsky-Moriya (DM) interaction and an additional c-axis
anisotropy.~\cite{lumsden} The exchange constant is small because
magnetic (S=1/2) V$^{4+}$-O$_{5}$ pyramids are separated by
nonmagnetic (S=0) V$^{5+}$-O$_4$ tetrahedra. Interlayer couplings
and anisotropies lead to an antiferromagnetic ordering at $T_N=4$~K,
showing unusual field induced spin reorientations.
Magnetization and neutron diffraction studies show
continuous spin reorientation in a basal-plane
magnetic field.~\cite{lumsden} Theoretical works suggest novel
magnetic structures, which consist of chiral helices with rotation
of staggered magnetization and oscillations of the total magnetization.~\cite{bogdanov,chernyshev}
Inelastic neutron scattering studies reveal an unexpected mode splitting of a spin wave near the ($\pi/2$,$\pi/2$) zone boundary.~\cite{lumsden06}

Thermal conductivity measurements give a hint of a local symmetry breaking as indicated by anomalies around 110~K.~\cite{sales} Optical spectroscopy measurements confirm a successive structural transition:
a distortion of the apical oxygen of the VO$_5$ square pyramids at about 110~K and a weak basal plane distortion around 60~K.~\cite{Withers,choi} Recent crystallographic measurements unveil a symmetry reduction to space groups {\it P4$_2$bc } or {\it P4nc} at 115~K.~\cite{chakoumakos}

The most remarkable feature is the observation of magnetodielectric effects by magneto-optical experiments:
(i) a redshift of the 1~eV excitation with a vibration
fine structure of the frequency 55~cm$^{-1}$, and
(ii) a narrowing of the 2.3~eV excitation
with increasing magnetic field.~\cite{Rai} The field-induced changes in the optical excitations were attributed to local distortions of the VO$_5$ square pyramids with
external magnetic field. All these seem to suggest that a soft lattice is coupled
to a spin system but clear spectroscopic evidence is missing. Raman spectroscopy can
serve as an experimental choice because it is extremely sensitive to
both lattice anomalies and low-lying spin excitations.

In this paper, we report on the observation of a soft mode
in A$_1$ symmetry and the two-magnon excitations in B$_1$ symmetry
which show a similar energy scale and temperature dependence.
The two-magnon continuum exhibits an unexpected double-peak lineshape.
These feature are interpreted as resulting from the coupling of
spin and lattice modes at finite momentum.

\section{EXPERIMENTAL DETAILS}
Single crystals of K$_2$V$_3$O$_8$ were grown in a platinum
crucible by the flux method. Samples with typical dimensions of $4\times 4\times
0.1 $~mm$^{3}$ were used for Raman scattering experiments. The
Raman spectra were measured in a continuous helium flow optical
cryostat by varying temperature from 1.5 K to room temperature.
All spectra were taken in a quasi-backscattering geometry with the
excitation line $\lambda= 514.5$ of an Ar$^{+}$ laser, a DILOR-XY
spectrometer, and a nitrogen cooled charge-coupled device detector.
The laser power of 6~mW was focused to a 100~$\mu$m diameter spot on the crystal.

\section{RESULTS AND DISCUSSION}

\subsection{Lattice instabilities}
%%%%%%%%%%%%%%%%%%%%%%%%%%%%%%%%%%%%%%%%%%%%%%%%%%%%%%%%%%%%%
%Figure 1
%%%%%%%%%%%%%%%%%%%%%%%%%%%%%%%%%%%%%%%%%%%%%%%%%%%%%%%%%%%%%
\begin{figure}[tbp]
\linespread{1}
\par
\includegraphics[width=8cm]{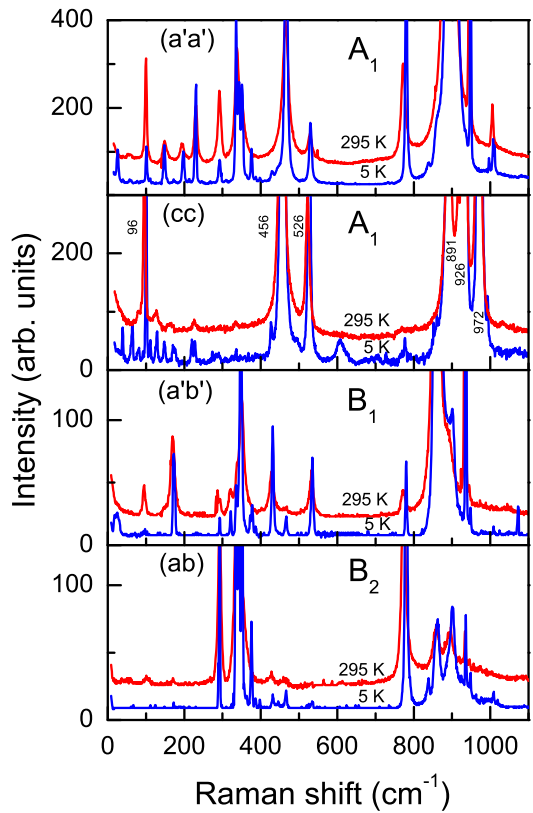}
\par
\caption{(Color online) Raman spectra of K$_2$V$_3$O$_8$ in four
        different (a$^{'}$a$^{'}$), (cc),
(a$^{'}$b$^{'}$), and (ab) polarizations
 at 5 K and 295 K.} \label{fig1}
\end{figure}

Before discussing Raman spectra we  present a factor group
analysis of the expected phonon modes through the structural transition.
At room temperature K$_2$V$_3$O$_8$ has a space group {\it P4bm}
and two formula units (Z=2). The factor group analysis yields
the following irreducible representations:

\begin{eqnarray}
\Gamma_{total} & = & 12A_{1}(a^{'}a^{'},cc, {\bf E\|c}) +
8A_{2} + 6B_{1}(a^{'}b^{'})\nonumber \\
 & + & 10B_{2}(ab)+ 21 E(ac,bc, {\bf E\|a},{\bf E\|b}).
\end{eqnarray}
Here A$_2$+ 2E modes are acoustic phonon modes.

For temperatures below 115~K a structural transition to space groups {\it P4$_2$bc } or {\it P4nc}
takes place.
For the {\it P4$_2$bc } space group we have
the following $\Gamma$-point vibrational modes:
\begin{eqnarray}
\Gamma_{total} & = & 15A_{1}(a^{'}a^{'},cc, {\bf E\|c}) + 15A_{2} +  15B_{1}(a^{'}b^{'})\nonumber \\
 & + &  15B_{2}(ab) + 36 E(ac,bc, {\bf E\|a},{\bf E\|b}).
\end{eqnarray}

For space group {\it P4nc} the total irreducible
representations are given as follows:
\begin{eqnarray}
\Gamma_{total} & = & 15A_{1}(a^{'}a^{'},cc, {\bf E\|c}) + 15A_{2} + 13B_{1}(a^{'}b^{'}) \nonumber\\
 & + &  13B_{2}(ab) + 32 E(ac,bc, {\bf E\|a},{\bf E\|b}).
\end{eqnarray}

In order to discriminate the phonon anomalies induced by the structural phase transition,
the Raman spectra were compared at 5 K and 295 K
in (a$^{'}$a$^{'}$), (cc), (a$^{'}$b$^{'}$), and (ab) polarizations, as
shown in Fig.~1. Here $a^{'}= a + b$ and $b^{'}= a - b$ are rotated
by $45^{\circ}$ from the respective $a$  and $b$ axis.
The spectra for the (a'a') and (cc) polarizations consist of A$_1$ symmetry modes.
The crossed (a'b') and (ab) polarizations correspond to the B$_1$ and B$_2$ mode,
respectively. The E modes which contribute interplane, crossed polarizations could not be studied due to a very thin thickness of the crystal.

At room temperature we observe a total of the $19A_{1}+8B_{1}+8B_{2}$ modes.
The extra $7~A_{1}+2~B_{1}$ modes might be due to either local lattice distortions,
which activate the silent 7~A$_2$ modes or a leakage of a selection rule.
In contrast, the missing $B_{2}$ modes are due to a lack of phonon intensity
as well as to their overlap with larger intensity excitations. At T=3~K
we identify  $36A_{1}+14B_{1}+13B_{2}$. The number of the $A_{1}$ modes is much larger than the expected one for both {\it P4$_2$bc } and {\it P4nc}.
The observed $B_{1}$ and $B_{2}$ modes are not compatible with the space group
{\it P4nc}. This indicates that the real crystal symmetry at 3~K is lower than
{\it P4$_2$bc } and {\it P4nc} due to an additional base plane distortion at 60~K.
Detailed crystallographic characterizations are needed to determine the exact
crystal symmetry for temperatures below 60~K.

%%%%%%%%%%%%%%%%%%%%%%%%%%%%%%%%%%%%%%%%%%%%%%%%%%%%%%%%%%%%%
%Figure 2
%%%%%%%%%%%%%%%%%%%%%%%%%%%%%%%%%%%%%%%%%%%%%%%%%%%%%%%%%%%%%
\begin{figure}[tbp]
\linespread{1}
\par
\includegraphics[width=8cm]{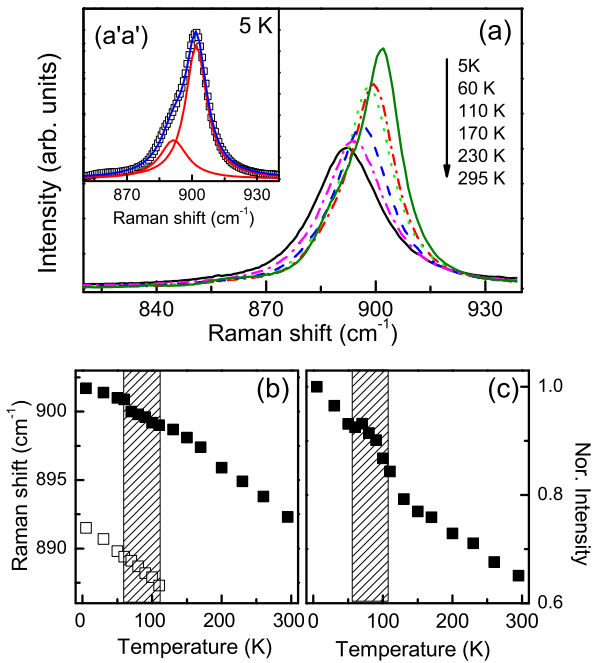}
\par
\caption{(a) Temperature dependence of the 891 cm$^{-1}$ mode.
Inset: A fit of the 891 cm$^{-1}$ mode to a sum of two Lorentzian
curves at 5~K.  (b) Temperature dependence of the peak position
of the 891 cm$^{-1}$ mode together with a new mode at a shoulder.
(c) Temperature dependence of the normalized integrated intensity
of the 891 cm$^{-1}$ mode.} \label{fig1}
\end{figure}

The vibrational modes are assigned in comparison with the experimental and theoretical Raman
works of the vanadium oxides AV$_2$O$_5$ (A=Li,Na,Cs,Mg,Ca), which has
a similar layered structure consisting of VO$_5$ square pyramids or VO$_4$ tetrahedra.
\cite{popovic00,popovic,spitaler} The high-frequency modes at $\omega > 500~\mbox{cm}^{-1}$ represent the
V-O stretching vibrations of the respective VO$_4$ tetrahedra and VO$_5$ pyramid.
The studied compound has two different V
sites and four different O sites: O1(bridging pyramid and tetrahedron),
O2(apical to tetrahedron), O3(bridging tetrahedra), and O4(apical to
pyramid). Since the bond lengths of V-O4 and V-O2 are 1.582~ and 1.628~{\AA}, respectively,
the highest frequency mode of 972~cm$^{-1}$ observed in the (cc) polarization
is the V-O4 stretching mode of the VO$_5$ square pyramid.
The 926~cm$^{-1}$ mode is the V-O2 stretching mode of  the VO$_4$ tetrahedra.
The 891~and 526~cm$^{-1}$ modes are the stretching modes of vanadium and
basal-plane oxygens. The phonon modes in the frequency range between 200 and 500 cm$^{-1}$
correspond to diverse O-V-O and V-O-V  bending vibrations. The low-frequency
$\omega < 200~\mbox{cm}^{-1}$ are assigned to K atom vibrations
and torsional and twisting motions of the square pyramids and tetrahedra.

In order to gain local information on lattice distortions through
the structural modulation, we highlight the phonon anomalies
in the respective low-, intermediate-, and high-frequency regimes.
Figure~2(a) shows the temperature dependence
of the 891 cm$^{-1}$ mode. At room temperature the peak consists
of a single Lorentzian line. Upon cooling through 110~K a new peak shows
up at the shoulder of the main peak. The temperature dependence of the frequency
exhibits a slight change in slope around 110~K and a weak jump
in the vicinity of 60~K. The normalized integrated intensity
also shows anomalies at the respective temperature.
The phonon frequency anomalies observed below 110~K are correlated with
the dramatic contraction of the $a$-lattice parameter at the respective
temperatures.~\cite{chakoumakos} Here we note that the infrared spectroscopy
also observed the splitting of the V-O4 stretching mode for temperatures below 110~K,
which was ascribed to the displacement of the apical oxygen from the apex
of the VO$_5$ square pyramid.~\cite{choi}

The Raman scattering intensity is given by the change of a crystal
polarizability with respect to the displacements of a normal mode.
Dipole matrix elements and interband transition energies
are two decisive factors in determining the intensity.~\cite{Sherman} Noticeably, the temperature
dependence of the 891-cm$^{-1}$ phonon intensity
resembles with that of the peak position of the V$^{4+}d\rightarrow d$ transition
[Compare Fig. 6 of Ref.~18 with Fig.~2(c)]. This suggests that
local distortions of the VO$_5$ square pyramids are closely related to electronic properties.

%%%%%%%%%%%%%%%%%%%%%%%%%%%%%%%%%%%%%%%%%%%%%%%%%%%%%%%%%%%%%
%Figure 3
%%%%%%%%%%%%%%%%%%%%%%%%%%%%%%%%%%%%%%%%%%%%%%%%%%%%%%%%%%%%%
\begin{figure}[t]
\linespread{1}
\par
\includegraphics[width=10cm]{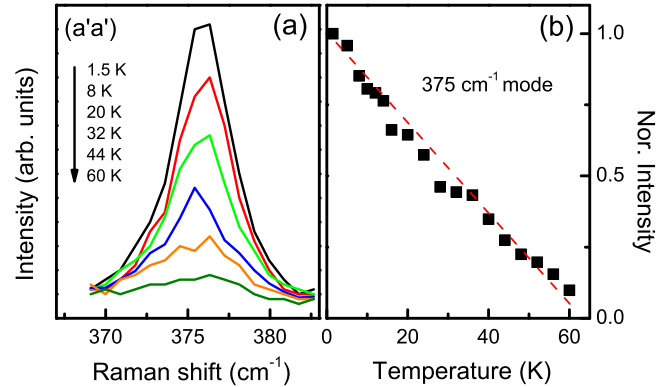}
\par
\caption{(a) Evolution of the 375 cm$^{-1}$ peak which appears
for temperatures below 60~K. (b) Temperature dependence of
the normalized integrated intensity of the 375 cm$^{-1}$ peak.
The dashed line is guide for the eyes. } \label{fig1}
\end{figure}

Figure~3 shows the intermediate-frequency mode  at
376 cm$^{-1}$, which is induced by the structural modulations
for temperatures below 60~K.
Since the new phonon mode corresponds to O-V-O bending vibrations,
it results from the basal plane distortions.
With decreasing temperature the intensity increases almost
linearly without saturation [see Fig.~3(b)]. This indicates that
the base plane modulations grow continuously all the way
down to the base temperature.

%%%%%%%%%%%%%%%%%%%%%%%%%%%%%%%%%%%%%%%%%%%%%%%%%%%%%%%%%%%%%
%Figure 4
%%%%%%%%%%%%%%%%%%%%%%%%%%%%%%%%%%%%%%%%%%%%%%%%%%%%%%%%%%%%%
\begin{figure}[t]
\linespread{1}
\par
\includegraphics[width=8cm]{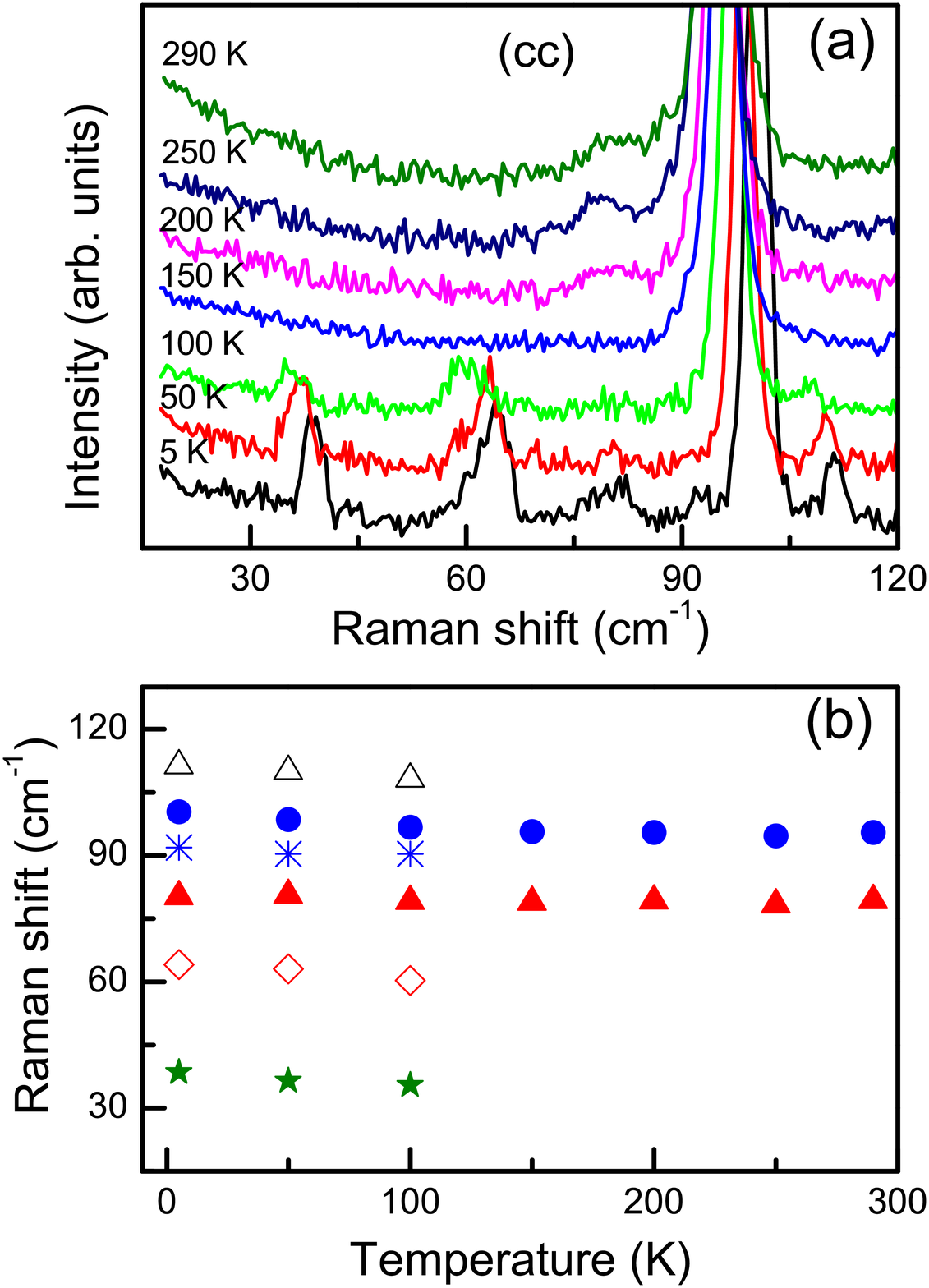}
\par
\caption{(Color online) (a) Temperature dependence of the low-frequency
Raman spectra in (cc) polarization. (b) Peak splitting of the 95 cm$^{-1}$ mode
as a function of temperature.} \label{fig1}
\end{figure}

The most striking lattice anomalies are observed in the low-energy
phonon modes in (cc) polarization as displayed in Fig.~4. The 95~cm$^{-1}$ mode
shows a splitting with decreasing temperature through 110~K.
It is remarkable that the infrared-active mode at 55~cm$^{-1}$ also exhibits
a similar splitting at the respective temperature.~\cite{Rai}
Since these low-frequency modes are associated with
the torsional and twisting motions of the square pyramidal and tetrahedral
building blocks, the structural modulations are a mixture of out-of-plane bending
and rotations of the vanadium oxide polyhedra.  We can summarize the
evolution of the structural modulations as follows. The rotational modulations of the vanadium oxide polyhedra start around 110~K and then the base plane distortions follow at 60~K. The hardening by 4-5~cm$^{-1}$ indicates that the twisting degree of the V pyramids and the tetrahedra
increases substantially upon cooling from 110~K.

\subsection{A soft mode and magnetic excitations}

%%%%%%%%%%%%%%%%%%%%%%%%%%%%%%%%%%%%%%%%%%%%%%%%%%%%%%%%%%%%%
%Figure 5
%%%%%%%%%%%%%%%%%%%%%%%%%%%%%%%%%%%%%%%%%%%%%%%%%%%%%%%%%%%%%
\begin{figure}[t]
\linespread{1}
\par
\includegraphics[width=9cm]{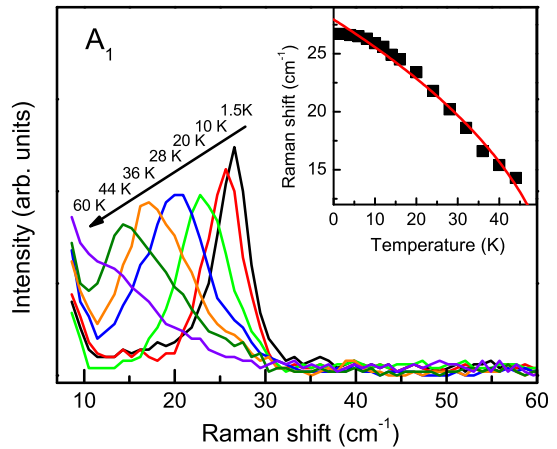}
\par
\caption{(Color online) Evolution of the 26-cm$^{-1}$ soft mode
in A$_1$ symmetry which is sensitive to the structural transition at 60~K.
Inset: Temperature dependence of the soft mode
frequency. The solid line is a fit to a power-law. See the text for details.} \label{fig1}
\end{figure}

We turn to more direct evidence for a soft lattice.
In second-order phase transitions
there will always be a zone-center
A$_{1}$ component of the soft mode below the critical temperature.~\cite{Worlock}
 Indeed, we observe the sharp peak
at 26~cm$^{-1}$ in the A$_1$ symmetry as shown in Fig.~5.
Upon approaching the phase transition temperature of 60~K the 26~cm$^{-1}$ mode
softens and then evolves into  quasielastic scattering.
We find no evidence for the presence of
a Raman-active soft mode related to the structural phase transition at~110 K.
The observed soft mode is assigned to
the torsional mode that flexes the VO$_5$ square pyramids out of the plane, which could not be detected by optical spectroscopy.~\cite{Rai} Here we note that the layer compound SrCu$_2$(BO$_3$) has a soft mode at 62~cm$^{-1}$ as an in-phase motion of almost all ions along an out-of-plane direction.~\cite{KYC} This soft
mode is a consequence of a continuous corrugation of the Cu-BO$_3$ layer. This analogy can be taken as further evidence for the basal plane distortions.

The inset of Fig.~5 displays the temperature dependence of the peak position of
the soft mode in a temperature range of 1.5 - 45~K. It is well described by a power law, i.e. $\omega = A  \left| T_C-T \right|^{\beta}$, with $A = 4.7$ cm$^{-1}$/K , $T_C = 55$ K
and $\beta = 0.448$. The observed critical exponent is close to a mean-field value of $\beta = 1/2$. For SrCu$_2$(BO$_3$) the soft mode undergoes an additional softening toward low temperatures due to spin-phonon coupling.~\cite{KYC}  This anomaly is not self-evident for the case of K$_2$V$_3$O$_8$ although the small discrepancy between
the experimental data and the fitted curve might have an origin of spin-phonon coupling. This is not surprising if the long V$^{4+}$-O-O-V$^{4+}$ superexchange paths are taken into account.

%%%%%%%%%%%%%%%%%%%%%%%%%%%%%%%%%%%%%%%%%%%%%%%%%%%%%%%%%%%%%
%Figure 6
%%%%%%%%%%%%%%%%%%%%%%%%%%%%%%%%%%%%%%%%%%%%%%%%%%%%%%%%%%%%%
\begin{figure}[t]
\linespread{1}
\par
\includegraphics[width=8cm]{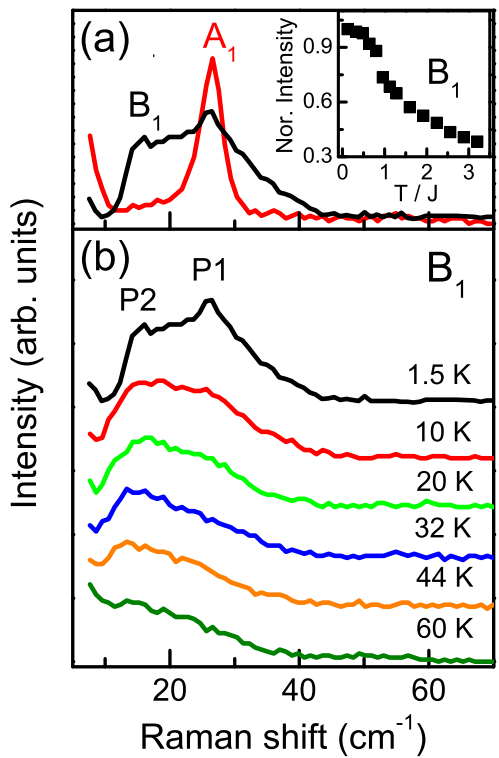}
\par
\caption{(Color online) (a) Comparison of the soft mode seen in the A$_1$ symmetry
and the two-magnon continuum observed in the B$_1$ symmetry.
Inset: Temperature dependence of the integrated intensity of
two-magnon scattering in a temperature scale of $J$.
(b) Temperature dependence of two-magnon scattering in
the  B$_1$ symmetry.} \label{fig1}
\end{figure}

We now switch to the low-energy spin excitation observed in
($a^{'}b^{'}$) polarization. Figure 6(a) compares the magnetic continuum  in
the B$_1$ symmetry with the soft mode in the  A$_1$ symmetry.
The continuum exhibits a double peak structure
extending from 10 cm$^{-1}$ up to $ 44$ cm$^{-1}$. The primary
peak (P1) is centered at 26 cm$^{-1}$ while the secondary peak (P2)
lies at 16~cm$^{-1}$. The primary peak position coincides with the soft phonon mode.
The magnetic continuum is ascribed to two-magnon excitations
by considering the energy scale of the spectral weight and the selection rule (see below).

In two-dimensional (2D) antiferromagnets, two-magnon scattering can be probed
as a double spin-flip process via the exchange light scattering.~\cite{Fleury}
A dominant contribution comes from a symmetry-allowed B$_{1}$
geometry.~\cite{KYC08} This selection rule strictly holds for our case.
The density of states of a two-magnon spectrum
is given by $\rho_{2}(\omega)=\sum_{k}\delta (\omega
-2\omega_{k})$, where $\omega_{k}$ is a one-magnon dispersion.
From the upper cut-off frequency of the two-magnon spectrum, we can determine
the magnon energy at the zone boundary. The obtained value of 2.8~meV($\approx
22.4$~cm$^{-1}$) agrees perfectly with the neutron scattering result (compare to
Fig.~12 of Ref.~\cite{lumsden06}).

In addition, we can estimate the exchange coupling constant between vanadium spins
by resorting to the upper cut-off relation, that is, $2Z_{c}JzS=4.632~J$, where $J$ is the exchange integral, $z$ is the number of nearest neighboring spins,
and $Z_{c}=1.158$ is a quantum correction to a classical value.~\cite{Igarashi,Singh}  We obtain $J\approx 13.6 K$.
This is in reasonable agreement with the value of $J=12.8~K$ extracted by inelastic
neutron scattering.~\cite{lumsden06} The small overestimation is due to the presence
of interlayer interactions and DM interactions. Here we note that a conventional way
of determining $J$ from the two-magnon continuum is based on the peak energy, which
corresponds to  $\omega_{p}\sim 2.7 J$.~\cite{cottam} However, this method is not
unambiguously applicable to K$_2$V$_3$O$_8$  since there is no well-defined single peak unlike other antiferromangets.~\cite{KYC08,cottam}

Figure 6(b) shows the temperature dependence of the two-magnon scattering.
The spectrum persists to T=60~K in the form of short-range magnetic fluctuations.
This temperature amounts to 15~T$_{N}$(= $5J$). The survival of
the zone boundary magnon to exceptionally high temperatures in an energy scale of $T_N$ reflects the fact that in 2D S=1/2 quantum spin systems magnetic excitations are scaled by an exchange coupling constant $J$ while
the classical N\'{e}el ordering is caused by interplane interactions.
Although the spectral weight of the two-magnon profile matches well with
the spin wave theory, a careful inspection of the continuum reveals
some anomalies.

First, the relative intensity falls off strongly with an upturn
about J [see the inset of Fig.~6(a)]. This behavior is contrasted
by other antiferromagnets, which show an increase of a total scattering
intensity or a moderate decrease in a high-temperature limit.~\cite{KYC08,cottam}

Second, the lineshape shows the double-peak structure. Since two-magnon scattering is dominated by zone boundary magnons, it might be due to the zone boundary
mode splitting. Indeed, inelastic neutron scattering identifies the split modes at the zone boundary centered
around $21~\mbox{cm}^{-1}$ with an energy difference of $4~\mbox{cm}^{-1}$.~\cite{lumsden06}
Since Raman spectroscopy probes magnetic excitations as the two-magnon continuum,
the energy difference between the two peaks is expected to double that of the neutron split modes.
The observed separation energy of $10~\mbox{cm}^{-1}$ is slightly larger than twice that of
the neutron split modes. Its possible reason is discussed below.

%%%%%%%%%%%%%%%%%%%%%%%%%%%%%%%%%%%%%%%%%%%%%%%%%%%%%%%%%%%%%
%Figure 7
%%%%%%%%%%%%%%%%%%%%%%%%%%%%%%%%%%%%%%%%%%%%%%%%%%%%%%%%%%%%%
\begin{figure}[t]
\linespread{1}
\par
\includegraphics[width=7cm]{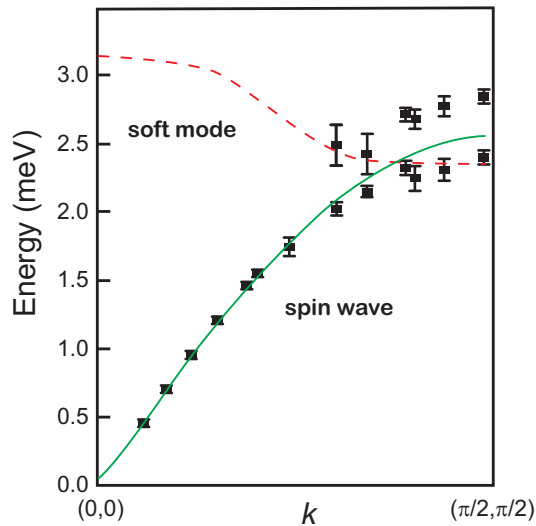}
\par
\caption{(Color online) A sketch of coupling of a soft mode to
a spin wave near the zone boundary. The spin wave data and their fit to
linear spin-wave theory (solid line) are reproduced from Fig.~12 of Ref.~\cite{lumsden06}.
The dashed line represents a typical optical dispersion of the soft mode,
which is based on the observed $\Gamma$-point energy of 26 cm$^{-1}$.  } \label{fig1}
\end{figure}

In order to explain the observed split modes Lumsden {\it et al.}~\cite{lumsden06} have discussed several possible scenarios, for example, magnon-phonon interaction, structural distortions, lattice disorders, multimagnon scatterings, and orbital degrees of freedom. However, no definite answer can be given. Based on the two-magnon lineshape and temperature dependence, we propose a coupling of a soft mode to magnetic excitations as an origin. The coincidence of the energy scales and temperature dependence makes it difficult to differentiate two modes using experimental techniques different from  Raman spectroscopy. Thanks to the Raman selection rules, however, we are able to separate them from  each other.

Here we remind the reader that there are soft optical and acoustic modes in structural phase transitions.
The former can be observed by Raman spectroscopy while the latter can be detected
by Brillouin  scattering techniques. Although the dispersion of the soft mode cannot
be directly determined by Raman scattering experiments, we can sketch a typical profile
by exploiting the fact that the soft optical mode has quite flat dispersion
and has a $\Gamma$-point energy of 26 cm$^{-1}$. The resulting dispersion curve
is illustrated schematically as the dashed line in Fig.~7. The soft phonon is presented together
with the spin wave obtained by inelastic neutron scattering experiments
(squares symbols and solid line in Fig.~7). Judging from the comparable energy scale of the soft mode at the (0,0) point and the magnon at the ($\pi/2,\pi/2$) point, it is apparent that the soft mode crosses the magnon near the zone boundary.
This provides a compelling explanation on the split modes. Since a mixing of the spin and phonon mode occurs near the zone boundary, a coupled spin-phonon character will be manifested
at finite momentum. Therefore, we expect no anomalous feature for the $\Gamma$-point soft mode as Fig.~5 shows. However, the lineshape and temperature dependence of the two-magnon scattering will be different from conventional antiferromagnets. As Fig.~6 shows, the spectral form and the scattering intensity reflect a hybrid nature.

Last, we  detail the anomalous double-peak feature in relation to the neutron split modes.
The zone boundary magnons
have energies of 19  and 23 cm$^{-1}$. The corresponding two-magnon density
of states has the upper cut-off energies at 38  and 46 cm$^{-1}$.
This energy  corresponds to $4.632~J$.~\cite{Igarashi,Singh} Due to magnon-magnon interactions
the two-magon peak is shifted down to $2.7~J$ for 2D antiferromagnets.~\cite{cottam} This relation
enables us to estimate the peak energies of the two-magnon
spectrum as 22  and 27 cm$^{-1}$.
Compared to the observed peak energies of 16 cm$^{-1}$ (P2) and 26 cm$^{-1}$ (P1),
only the lower-energy peak shows an appreciable energy difference by 6~cm$^{-1}$.
The discrepancy suggests that there are additional softening channels,
specific to the lower-energy branch. This might be due to the fact that
the lower branch of the hybrid modes has a more phonon character.

\section{CONCLUSION}

To conclude, we have presented Raman scattering studies of S=1/2 2D antiferromagnet K$_2$V$_3$O$_8$. Owing to magnetodielectric couplings, this system shows marked
anomalies in phonon and magnetic excitations.  Successive structural phase transitions at 110~K and 60~K are confirmed by pronounced phonon anomalies in frequencies, intensities, and numbers.  Further, for temperatures below 60~K we observe a soft mode at 26~cm$^{-1}$ in A$_1$ symmetry and two-magnon excitations in B$_1$ symmetry. The two-magnon continuum shows an unexpected double-peak structure and a pronounced temperature dependence. This is interpreted in terms of a mixing of spin and lattice modes near the zone boundary. Our result suggests that for magnetodielectric materials a mixing of phonon-magnetic modes occurs at finite momentum.

\section*{Acknowledgments}
This work was supported by NTH and DFG. K.Y.C. acknowledges financial support from the Humboldt Foundation and the NRF of Korea, Grant No. 2009-0093817.

\end{document}